\documentclass[aps,prl,showpacs,notitlepage,floatfix, twocolumn,
superscriptaddress]{revtex4-1}
\usepackage{bbm}
\usepackage{mathrsfs}
\usepackage{epsfig}
\usepackage{graphicx}
\usepackage{amsfonts}
\usepackage[figuresright]{rotating}
\usepackage{amssymb}
\usepackage{amsmath}
\usepackage{dcolumn}
\usepackage{bm}
\usepackage{braket}
\usepackage[colorlinks,linkcolor=blue,anchorcolor=blue,citecolor=blue,urlcolor=blue]{hyperref}
\usepackage[titletoc]{appendix}
\usepackage[normalem]{ulem}

\def\be{\begin{equation}}
\def\ee{\end{equation}}
\def\bea{\begin{equation}\begin{aligned}}
\def\eea{\end{aligned}\end{equation}}

\def \tr {\text{tr}}

\begin{document}
\title{Butterfly effect in interacting Aubry-Andre model: thermalization, slow scrambling, and many-body localization}

\author{Shenglong Xu}
\affiliation{Condensed Matter Theory Center and Joint Quantum Institute, Department of Physics,  University of Maryland, College Park, MD 20742, USA}

\author{Xiao Li}
\affiliation{Condensed Matter Theory Center and Joint Quantum Institute, Department of Physics,  University of Maryland, College Park, MD 20742, USA}
\affiliation{Department of Physics, City University of Hong Kong, Kowloon, Hong Kong, China}

\author{Yi-Ting Hsu}
\affiliation{Condensed Matter Theory Center and Joint Quantum Institute, Department of Physics,  University of Maryland, College Park, MD 20742, USA}

\author{Brian Swingle}
\affiliation{Condensed Matter Theory Center, Maryland Center for Fundamental Physics, Joint Center for Quantum Information and Computer Science,
and Department of Physics, University of Maryland, College Park, MD 20742, USA}

\author{S. Das Sarma}
\affiliation{Condensed Matter Theory Center and Joint Quantum Institute, Department of Physics,  University of Maryland, College Park, MD 20742, USA}

\begin{abstract}
The many-body localization transition in quasiperiodic systems has been extensively studied in recent ultracold atom experiments. At intermediate quasiperiodic potential strength, a surprising Griffiths-like regime with slow dynamics appears in the absence of random disorder and mobility edges. In this work, we study the interacting Aubry-Andre model, a prototype quasiperiodic system, as a function of incommensurate potential strength using a novel dynamical measure, information scrambling, in a large system of 200 lattice sites. Between the thermal phase and the many-body localized phase, we find an intermediate dynamical phase where the butterfly velocity is zero and information spreads in space as a power-law in time. This is in contrast to the ballistic spreading in the thermal phase and logarithmic spreading in the localized phase. We further investigate the entanglement structure of the many-body eigenstates in the intermediate phase and find strong fluctuations in eigenstate entanglement entropy within a given energy window, which is inconsistent with the eigenstate thermalization hypothesis. Machine-learning on the entanglement spectrum also reaches the same conclusion. Our large-scale simulations suggest that the intermediate phase with vanishing butterfly velocity could be responsible for the slow dynamics seen in recent experiments.\end{abstract}
\maketitle

\textit{\color{blue}Introduction.}---Much recent experimental and theoretical effort has been devoted to studying the conditions under which an isolated quantum system comes to effective thermal equilibrium. While thermalization is expected in many cases, an interacting many-body system may fail to thermalize due to strong quenched disorder, leading to a breakdown of the eigenstate thermalization hypothesis (ETH)~\cite{ETH1,ETH2,Rigol2008}. This phenomenon, now known as many-body localization (MBL)~\cite{MBL01,MBL02,MBL03,MBL04,MBL05,MBL06,Iyer2013Many-body,MBL08,MBL09,MBL10,MBL11,MBL12,MBL13,Potter2015,NEM_PRL_Mukerjee,XiaoPeng_PRL,XiaoPeng_PRB,MBL16,MBL14,Imbrie2016,
MBL17,MBL15,MBL18,NEMdisorderspin,NEM_JJunction,NEM_DensityPropagator,Parameswaran2018}, has greatly expanded our understanding of out-of-equilibrium dynamics in interacting quantum many-body systems.

MBL can also occur in deterministic quasiperiodic systems without any quenched disorder~\cite{Iyer2013Many-body,Khemani2017}, with most existing MBL experiments in ultracold atoms,  focusing on a special type of quasiperiodic system called the interacting Aubry-Andre (AA) model~\cite{aubry1980analyticity}, given in Eq.~\eqref{eq:H} below.
In the non-interacting limit, the AA model exhibits a singe localization transition at a quasiperiodic potential strength $\lambda_c = 1$; all single-particle eigenstates are extended (localized) when $\lambda$ is smaller (larger) than $\lambda_c$. In particular, the non-interacting AA model thus does not possess a single-particle mobility edge, by virtue of a special self-duality property, in contrast to generalized AA (GAA) models which manifest moblity edges~\cite{GAASPME,SPME_GAA_Theory_2017,SPME_GAA_Experiment_2018,Li2018}. With interactions, exact diagonalization (ED) studies in a small system predict that the MBL transition occurs at a potential strength considerably larger than $\lambda_c$~\cite{Iyer2013Many-body}.

Recent experiments~\cite{Schreiber_2015,Lueschen2017,Kohlert2018} have observed an MBL transition in the AA model at a critical potential strength similar to the numerical predictions.
Surprisingly, before the MBL transition there exists a substantial range of $\lambda$ in which the memory of the initial state relaxes very slowly, although the system is presumbly in a thermal phase~\cite{Lueschen2017,Kohlert2018}. Proposed explanations for this slow dynamics include local fluctuations in the initial state and atypical transition rates between single-particle states~\cite{Lueschen2017,BarLev2017,Lee2017}. However, our understanding of this regime of slow relaxation in the AA model is still rather incomplete, since it cannot arise from the usual Griffiths physics of rare spatial regions in a deterministic system. In this work, we study the structure and dynamics of entanglement, particularly the process of quantum information scrambling, to better understand the slow dynamics observed in experiments.

Scrambling \cite{Hayden2007, Sekino2008, Shenker2014, Hosur2016} describes the process whereby an initially localized perturbation spreads over the degrees of freedom of a complex quantum system. It can be quantified by the squared commutator,
\bea
C(r, t)=\braket {[W_0(t), V_r]^\dagger [W_0(t), V_r]}
\eea
where $W_0$ and $V_r$ are local operators. When expanded as a sum of correlation functions, one gets both time-ordered and out-of-time-order correlators (OTOC) \cite{Larkin1969}, so we will refer to $C$ as an OTOC. The physical picture is that the Heisenberg operator $W_0(t) = e^{i H t} W_0 e^{-i H t}$ expands in space as a function of time, so that $C(r,t)$ is close to zero when $V_r$ is far away from $W_0(t)$ and reaches $\mathcal{O}(1)$  when $V_r$ is within the support of $W_0(t)$.

In a local system, the speed at which an operator can expand is bounded by the Lieb-Robinson bound~\cite{Lieb1972}. The actual speed of operator expansion, as measured by the contours of constant $C$, is called the butterfly velocity $v_B$; it can be substantially less than the upper bound. In particular, disorder can impede operator growth. As a result, in disordered systems, the causal lightcone is sensitive to different dynamical phases~\cite{Swingle2016a, Chen2016a, Fan2017, Huang2017, Nahum2017, Sahu2018}: linear in the thermal phase, power-law like in the Griffiths phase and logarithmic in the MBL phase. Therefore, the OTOC offers a novel dynamical characterization of MBL systems and may shed light on the slow dynamics in the interacting AA model.

In this work, we carry out an extensive numerical study of the interacting AA model in order to understand possible origins of the slow dynamics observed experimentally. 
First, we calculate the OTOC using tensor network techniques in a large system~\cite{Xu2018} and find that there exists an intermediate dynamical phase between the thermal phase and the MBL phase, which we denote as the S phase.
This phase is characterized by a vanishing butterfly velocity and a power-law-like causal lightcone and is associated with slow relaxation of the initial state.
Second, we demonstrate using exact diagonalization (ED) method in a small system that the onset of the S phase is different from the usual MBL transition determined by entanglement entropy (EE) of the many-body energy eigenstates. We further show that the energy spectrum in the S phase contains a mixture of volume-law states and area-law states, and the latter may be responsible for the slow dynamics. Finally, we examine the entanglement spectrum (ES)~\cite{Haldane2008PRL} of energy eigenstates because the %ES encodes has proven
ES has been proven to be a sensitive probe of various dynamical phases~\cite{ESlevel_MBLTH,MLES_MBL_Titus,ML_ES_MBLTH,YangPRL2015,powerlawES,Hsu2018}.
We exploit the pattern recognition ability of machine-learning techniques~\cite{MelkoNphysIsing,van2017Learning,Chng2017Machine,Broecker2017Quantum} to study the complicated patterns inherent in the ES, a technique that has recently shown great promise for exploring dynamical phase diagrams~\cite{Hsu2018,ML_ES_MBLTH,MLES_MBL_Titus}.
The existence of three regimes naturally emerges from our machine learning results, and the three measures together robustly predict an intermediate distinct S phase in the interacting AA model at half filling. We believe that the S phase is responsible for experimental slow dynamics observed in the interacting AA model.

We emphasize that the S phase we report here is naively different from the intermediate phase observed in certain GAA models~\cite{NEM_PRL_Mukerjee,XiaoPeng_PRL,XiaoPeng_PRB,Hsu2018} or in systems with quenched disorder~\cite{NEMdisorderspin,NEM_JJunction,NEM_DensityPropagator}.
In the former case, the many-body intermediate phase may be attributed to the existence of a single-particle mobility edge in the noninteracting limit, while in the latter case rare region Griffiths physics may play a role. In contrast, neither mechanism is relevant in the AA model, so we think that the S phase arises purely from the interplay between the interaction and the incommensurate potential. 

\textit{\color{blue} The model.}---We consider the one-dimensional interacting Aubry-Andre (AA) model~\cite{aubry1980analyticity} with nearest neighbor interaction, described by the following Hamiltonian,
\bea
H=& -  J \sum \limits_{\braket{ij}} \left (c^\dagger_i c_j + h.c. \right ) +2\lambda \sum _i \cos (2\pi q i +\phi) n_i  \\
&+V\sum\limits_{\braket{ij}}n_i n_j,
\label{eq:H}
\eea
where $c_i$ is the fermion annihilation operator at site $i$, $n_i =c^\dagger_i c_i$ is the on-site particle number, and $\phi$ is a global phase in the potential. The hopping amplitude $J$ is set to be the energy unit. The on-site potential is incommensurate with an irrational wavenumber $q$.

%In the noninteracting limit ($V=0$), the system exhibits a single transition from the delocalized phase to the localized phase at a critical $\lambda_c=1$. 
We note that in the absence of interaction, the AA model has an inherent self-duality at $\lambda=1$ which plays a key role in its localization transition---mobility edges appear in the system when this self-duality is broken by any perturbation \cite{GAASPME, SPME_GAA_Theory_2017, SPME_GAA_Experiment_2018, Li2018}. 
Interaction effects lead to thermalization for small $\lambda$, implying that local observables measured in energy eigenstates are indistinguishable from those measured in the Gibbs ensemble at a temperature determined by the energy of the state. On the other hand, previous studies of the entanglement entropy and energy level statistics established that localization is stable against interaction for large $\lambda$~\cite{Iyer2013Many-body}, i.e. there exists a stable MBL phase. What is more intriguing is that slow power-law like relaxation has been found in numerics and experiments for $\lambda$ well below the MBL transition~\cite{Schreiber_2015, Lueschen2017, BarLev2017, Lee2017,Weidinger2018}.

To investigate the puzzling slow dynamics seen before localization, we study interaction effects in the AA model using scrambling as a diagnostic in a system with $L=200$ sites. This study is supplemented with an analysis of the entanglement structure of the many-body energy eigenstates in systems at half filling up to $L=18$ sites using ED methods. In the rest of this paper, we fix the irrational wavenumber $q=2/(\sqrt{5}+1)$ and the interaction strength $V=1$, and study the properties of the AA model as a function of $\lambda$.

\begin{figure*}
\includegraphics[height=0.35\textwidth, width=1.0\textwidth]{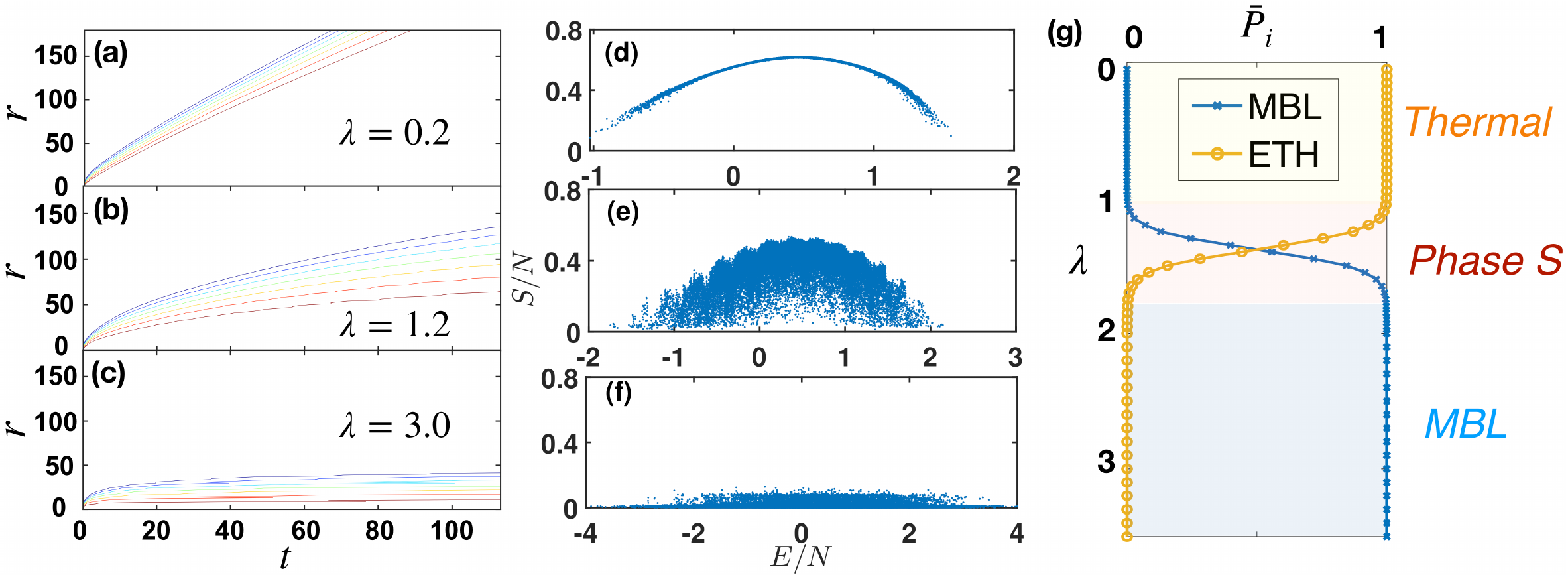}
\caption{The characteristic features of the three dynamical phases in the interacting AA model. (a)-(c) The contours of $\log (C(r,t))$ for three typical $\lambda$ representing each phase.  The shape of the causal lightcone is linear for the thermal phase (a), power-law like for the S phase (b) and logarithmic for the MBL phase (c).
(d)-(f) The von Neumann entropy of the many-body eigenstates as a function of energy in a system with $L=18$ sites at half-filling for the same $\lambda$ values as in (a)-(c). Here $N=L/2$ is the number of particles.
(g) The phase diagram produced by supervised machine learning of the many-body eigenstate ES from the middle energy spectrum. The classifier is trained by data from $\lambda$ deep in the thermal phase ($\lambda=0.2$) and MBL phase ($\lambda=3.0$), respectively.
$\bar{P}_i$ is the average confidence the classifier has to identify an input ES as belonging to phase $i=$ thermal or MBL over 2000 input data.}
%The von Neumann entropy shown in (d)-(f) and the input ES data used in (g) are produced by ED in a system with $L=18$ sites at half filling.}
\label{fig:regime}
\end{figure*}
%-------------------------------------------------------------------------------------------------------------------------------------

\textit{\color{blue}Three phases.}---We first calculate the following OTOC in the infinite-temperature Gibbs ensemble,
\bea
C(r, t)=\frac{1}{2^L}\tr \left ( [n_0(t),n_r(0)]^\dagger [n_0(t),n_r(0)] \right ) ,
\eea
using t-DMRG on the operator dynamics.
Due to the space-time structure of the entanglement entropy of the Heisenberg operator, long times can be studied ($\sim 150$ tunneling times) and the shape of the causal lightcone can be obtained accurately \cite{Xu2018}. In Fig.~\ref{fig:regime}(a)-(c), we present contour plots of $\log C(r,t)$ for three representative values of $\lambda$ for bond dimension $\chi=40$ (convergence with bond dimension has been checked). At small $\lambda=0.2$, the OTOC exhibits a linear lightcone [Fig.~\ref{fig:regime}(a)], indicating that the operator expands ballistically as expected for a thermal phase.
In the other limit where $\lambda=3.0$, the lightcone becomes logarithmic [Fig.~\ref{fig:regime}(c)], as expected for an MBL phase.
Interestingly, there exists an intermediate regime, where the causal light cone is power-law-like, shown in Fig.~\ref{fig:regime}(b) for a representative potential strength $\lambda=1.2$, which is below the nominal critical MBL value.
The phenomenon looks similar to that of the Griffiths phase in MBL systems with quenched disorder \cite{Nahum2017, Sahu2018}. A crucial difference is that, in the random disorder case, the power-law like causal lightcone only appears after disorder averaging and can be intuitively explained by the rare-region effects. For a single disorder realization, the rate of operator growth is significantly affected by specific disorder features and dramatically slows down in certain ``bottleneck'' regions. By contrast, in the interacting AA model, the power-law causal lightcone appears in a single realization of the incommensurate potential (the phase $\phi$ in  Eq.~\eqref{eq:H} is set to zero). Furthermore, we have checked that the shape of the lightcone does not depend on $\phi$. Therefore, rare regions in space do not play a role in this purely deterministic potential situation.

The appearance of the unexpected intermediate regime in the OTOC leads us to investigate the entanglement structure of the many-body energy eigenstates using ED methods.
In Fig.~\ref{fig:regime}(d)-(f), we plot the half-chain entanglement entropy (EE) $S=-\tr \rho\log\rho$ of the eigenstates as a function of energy in a system with $L=18$ sites, where $\rho$ is the half-chain reduced density matrix.
Consistent with the OTOC results, three quantitatively different entanglement regimes are observed, corresponding to the three dynamical phases we have identified.
In the small-$\lambda$ thermal phase, the EE density nicely collapses onto an energy-dependent curve, a hallmark of ETH;
in the opposite limit of the large-$\lambda$ localized phase, the EE of \textit{all} eigenstates are drastically reduced and exhibit area-law scaling;
in between, as shown in Fig.~\ref{fig:regime}(f), the many-body spectrum hosts a mixture of MBL-like and ETH-like eigenstates, indicating the S phase. 
In particular, while the volume-like behavior of EE is still visible from the outermost shape, a significant fraction of states clearly exhibit very low EE. We interpret this result as an indication of strong ETH violation but not complete localization. We speculate that this intermediate S-phase is causing the slow dynamics reported for the interacting AA model.

To further demonstrate the existence of the S-phase, we investigate how the ES of eigenstates evolve with $\lambda$ by constructing a neural network that classifies whether a given input ES is MBL-like or ETH-like. We generate the input ES data by ED. For each input ES, this classifier~\footnote{The network contains a hidden layer of $20$ sigmoid neurons.} produces two real numbers $P_{\text{MBL}}\in[0,1]$ and $P_{\text{thermal}}=1-P_{\text{MBL}}$, representing the confidence that the classifier identifies the input ES as belonging to MBL and thermal phases, respectively.
To obtain the phase diagram as a function of $\lambda$, we first train the network using ES data from $\lambda=0.2$ and $3.0$ for the thermal and MBL phase respectively.
We then feed the ES data from different $\lambda$'s to the trained classifier to obtain the average confidence $\bar{P}_{\text{MBL}}$ and $\bar{P}_{\text{thermal}}$ at each $\lambda$ [see Fig.~\ref{fig:regime}(g)].

We find that for $\lambda<0.95$ and $\lambda>1.9$, the network shows over $99.9\%$ confidence that the input ES data belong to thermal and MBL phases, respectively, as expected. Starting from $\lambda=0.95$, however, the gradual decrease of $\bar{P}_\text{MBL}$ suggests that the number of input ES that are ETH-like gradually decreases as $\lambda$ increases, with nearly all ES data becoming MBL-like around $\lambda=1.9$. The coexistence of MBL-like and ETH-like ES (from the middle of the energy spectrum) at a single $\lambda$ is consistent with what one would expect for the S phase, as indicated by the EE shown in Fig.~\ref{fig:regime} (e).

\begin{figure}[!]
\includegraphics[height=0.49\columnwidth, width=0.49\columnwidth]
{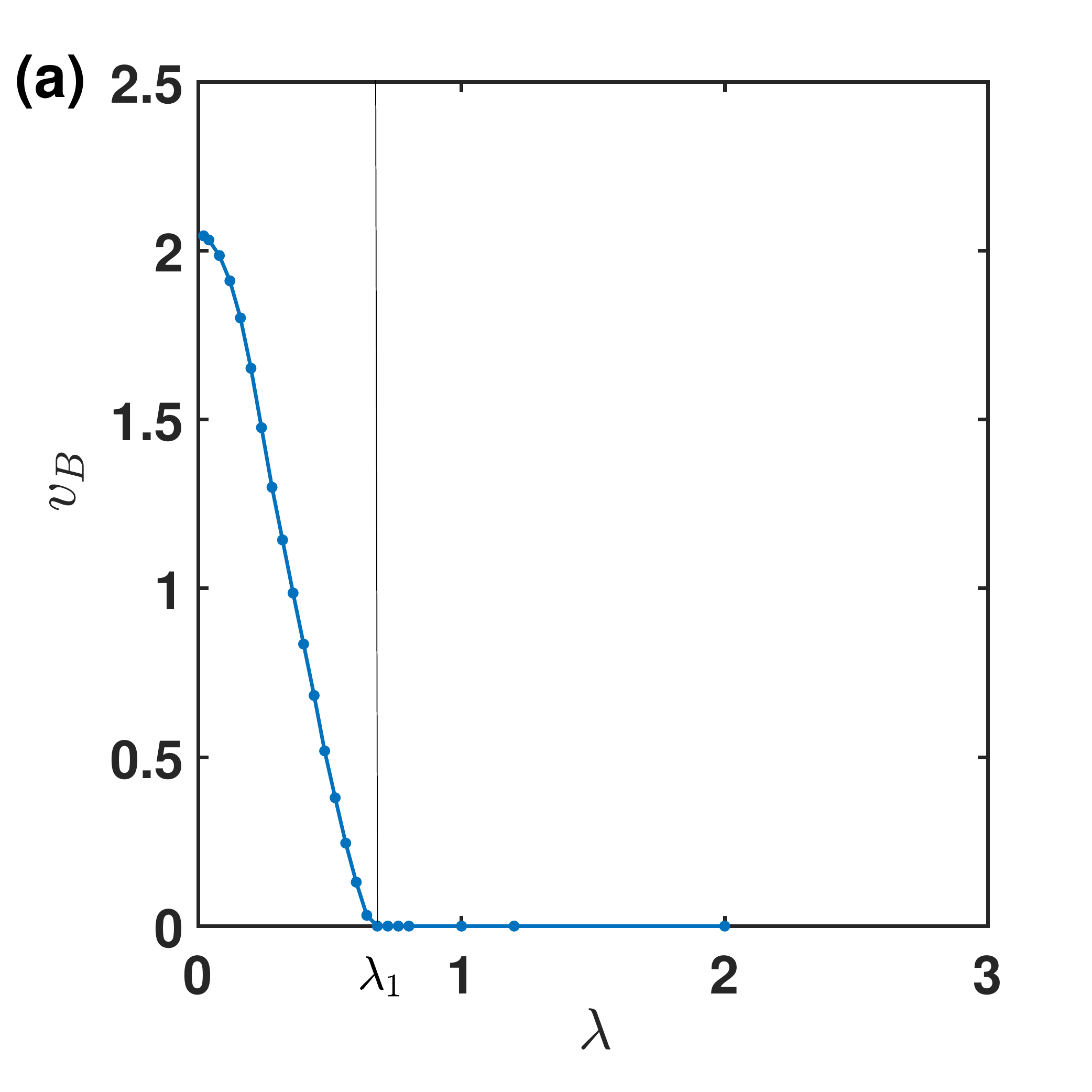}
\includegraphics[height=0.49\columnwidth, width=0.49\columnwidth]
{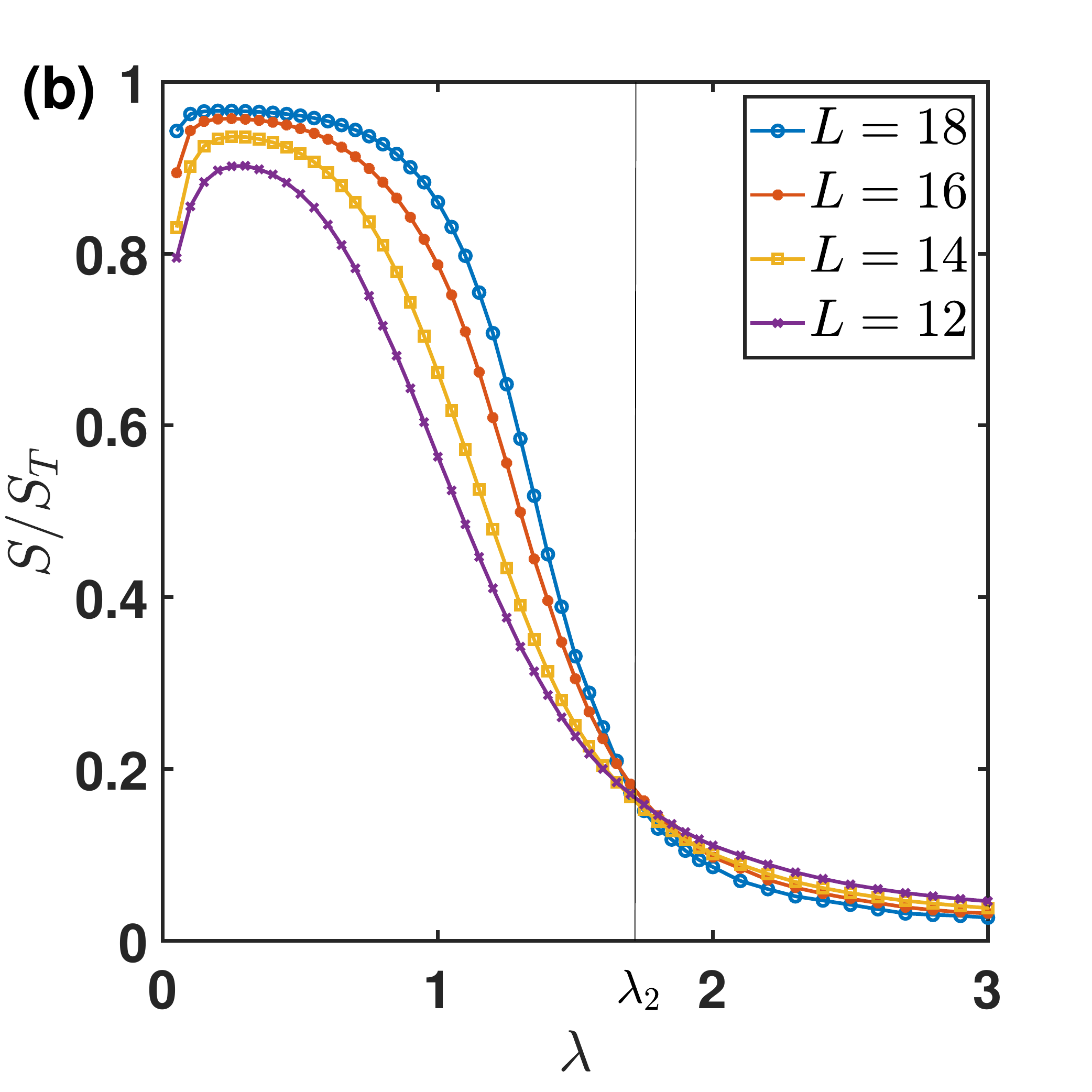}
\caption{The transitions among the three phases. (a) The butterfly velocity $v_B$ as a function of $\lambda$ in a system of $L=200$ sites. The transition between the thermal phase and the S phase is characterized by vanishing $v_B$ occurring at $\lambda \sim 0.7$. (b) Finite-size scaling of the eigenstate entanglement entropy with respect to the Page value $S_T$~\cite{Khemani2017} for $L$ varying from $12$ to $18$. The crossing point indicates the transition between the S phase and MBL phase at $\lambda \sim 1.7$.}
\label{fig:transition}
\end{figure}

\textit{\color{blue}Transitions.}---We have thus far identified three different dynamical regimes (small $\lambda$, thermal; large $\lambda$, MBL; in-between , S-phase) in the interacting AA model and described their characteristic features. Now we determine the two transition points. The first transition from the thermal phase to the S phase can be identified by the point where the causal lightcone changes from linear to power-law-like, which we denote as the slowing transition between thermal and S phases. The butterfly velocity $v_B$ provides a clean diagnostic of the phase transition between these two phases since $v_B$ is non-zero in the thermal phase and zero in the S phase. To extract $v_B$ from our data, we fit the early growth regime of OTOC ($-50 < \log C < -10$) to the form~\cite{Xu2018, Xu2018a, Khemani2018}
\bea
C(r,t)\sim \exp\left (-\alpha \frac{( r - v_B t)^{1+p}}{t^p}\right),
\eea
which applies to the region $r>v_B t$, and interpolates between the linear causal lightcone ($v_B > 0$)  and the power-law one ($v_B=0$) with $r \sim t ^ {p/(1+p)}$.
For non-interacting systems, it is expected that $v_B$ drops to zero at the single-particle localization transition, and we have checked that the velocity indeed drops to zero at $\lambda = \lambda_c=1$ when $V=0$, implying the non-existence of any intermediate S phase in the noninteracting AA model. However, for interacting systems, the slowing transition does not coincide with the MBL transition. The butterfly velocity of the interacting system is plotted in Fig.~\ref{fig:transition}(a); it dramatically decreases with increasing $\lambda$ and eventually vanishes at $\lambda_1 \sim 0.7$, signaling the thermal-to-S transition.

The second transition from the S phase to the MBL phase in the OTOC is characterized by a transition from a power-law lightcone to a logarithmic one.
However, using currently accessible space-time regions of the OTOC, it is difficult to precisely locate this transition point.
Instead, we perform the standard analysis of eigenstate EE in the middle of the energy spectrum, as shown in Fig.~\ref{fig:transition}(b), and the results indicate that the commonly accepted MBL transition happens at $\lambda_2 \sim 1.7$, much larger than $\lambda_1$ ($\sim 0.7$)  determined above from $v_B$ for the transition from the thermal phase to the S-phase.
Hence, interactions induce an intermediate dynamical phase of non-zero width ($0.7 \sim 1.7$) in $\lambda$ in sharp contrast with the non-interacting behavior.

%--------------------------------------------------------------------------------------------------------------------------------------
%-------------------------------------------------------------------------------------------------------------------------------------

\begin{figure}[!]
\includegraphics[height=0.49\columnwidth, width=0.49\columnwidth]
{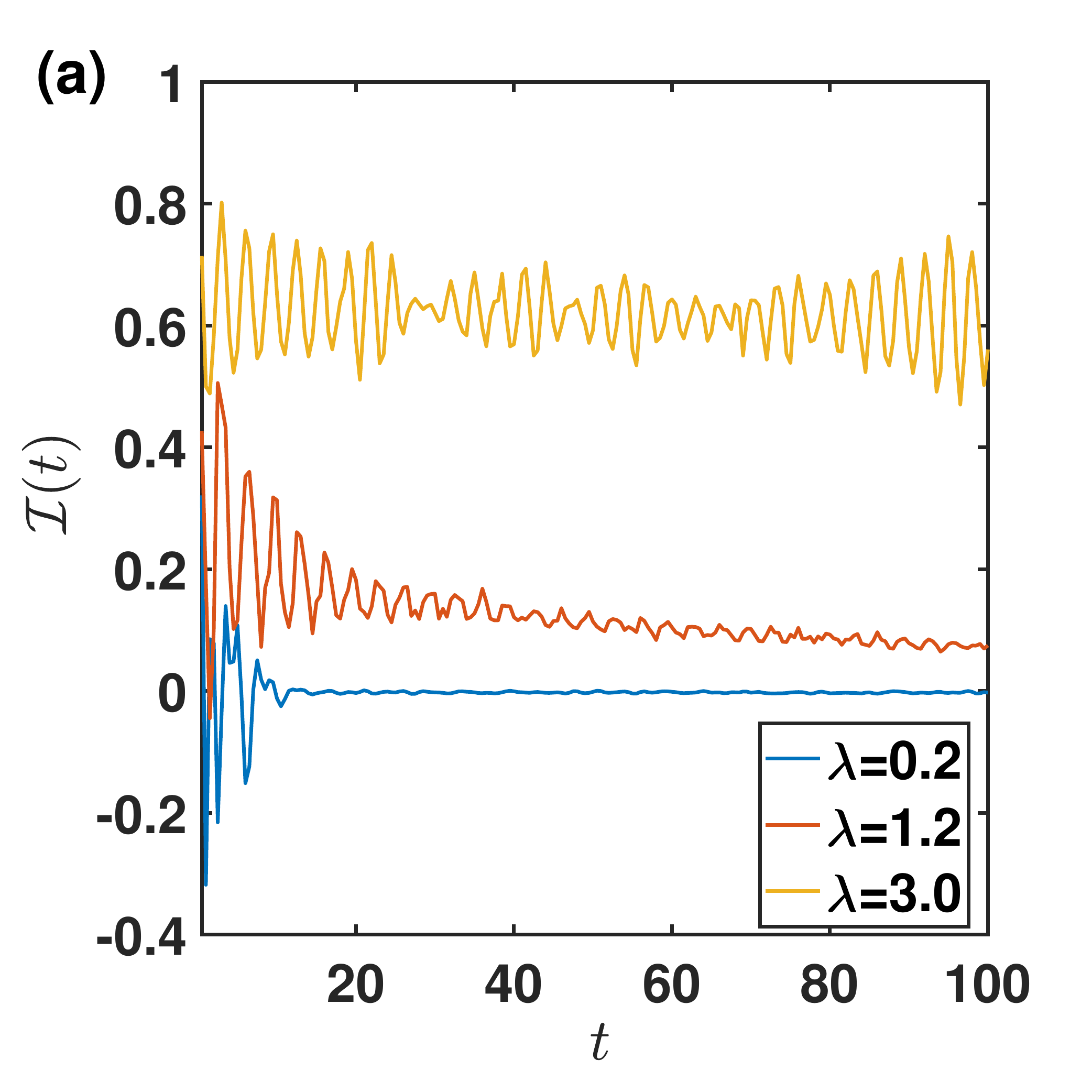}
\includegraphics[height=0.49\columnwidth, width=0.49\columnwidth]
{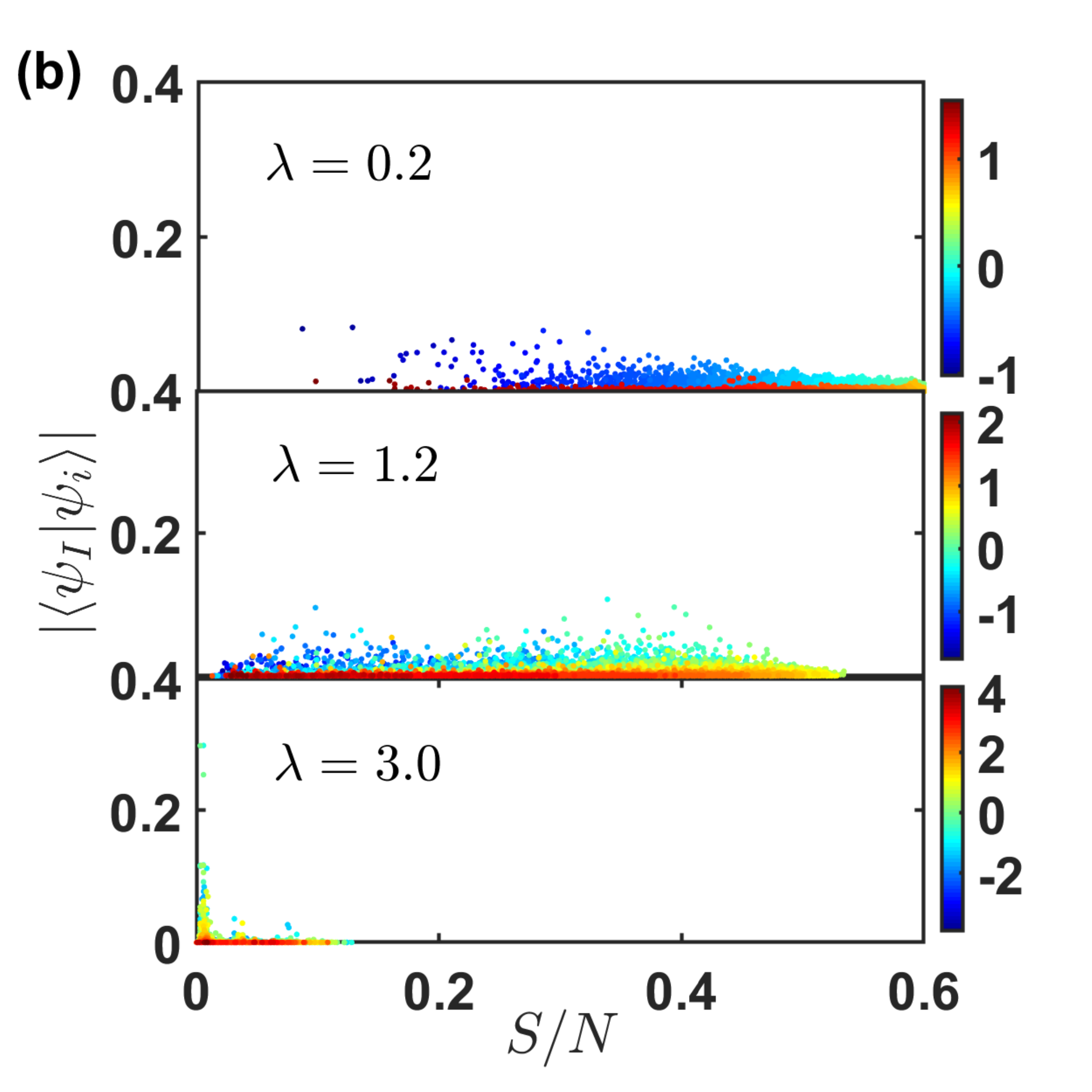}
\caption{\label{fig:slow} Slow dynamics. (a) Time evolution of the even-odd imbalance for three typical $\lambda$ values in a system of $L=100$ sites, exhibiting distinct asymptotic behaviors in the three dynamical phases.
(b) The overlap of the initial state $\ket{I}$ with all eigenstates as a function of the entanglement entropy density $S/L$ in a system of $L=18$ sites and $N=L/2$ particles. The color bar represents the energy density of the many-body eigenstates. }
\end{figure}

\textit{\color{blue}Slow dynamics.}---We speculate that the S phase is responsible for the slow dynamics of the interacting AA model recently observed in ultracold atom experiments \cite{Lueschen2017}.
The system is initialized in a density-wave state $\ket{\psi_I}$, where all the atoms occupy only odd lattice sites. Then the system is allowed to evolve in time and the even-odd imbalance $\mathcal{I}(t) = \sum_j (-1)^j n_j(t)$ is measured up to $\sim 100$ tunneling times. We stuided the imbalance using time-dependent variational principle in the matrix-product state manifold \cite{Haegeman2016, Leviatan2017} in a system of $L=100$ sites.
The results are plotted in Fig.~\ref{fig:slow}(a) for three typical values of $\lambda$ representing the three different phases we have studied (the results are converged with bond dimension $64$). We find that the imbalance exhibits three distinct behaviors on intermediate time scales ($10<t<100$)---exponential relaxation in the thermal phase ($\lambda=0.2$), freezing in the MBL phase ($\lambda=3.0$), and most importantly, slow decay in the S phase ($\lambda=1.2$), which is consistent with the experimental observations~\cite{Lueschen2017,Kohlert2018}, while even longer time behavior of the decay is not resolved yet \cite{BarLev2017, Weidinger2018}.
%--------------------------------------------------------------------------------------------------------------------------------------
%--------------------------------

As shown in Fig.~\ref{fig:regime}(d), in the S phase there are quite a few states with atypical low entanglement entropy in the middle of the spectrum.
Intuitively, these eigenstates should be responsible for the slow dynamics. To give evidence for this, we plot in Fig.~\ref{fig:slow}(b) the overlap between energy eigenstates and the initial density-wave state $\ket{\psi_I}$ as a function of the eigenstate EE, with the color map representing the energy density of the corresponding eigenstate. In the S phase, $\ket{\psi_I}$ has significant weight on non-thermal eigenstates, causing the slow relaxation of the initial state.

\textit{\color{blue}Conculsion.}---We have established the existence of a dynamically slow S phase in the interacting AA model, lying between the well-established thermal and MBL phases. The mechanism for the S phase in the AA model remains somewhat mysterious. Recent ultracold atom experiments~\cite{Kohlert2018} demonstrated that the AA model (without any single-particle-mobility-edge) and a generalized AA (GAA) model exhibit qualitatively similar dynamical properties as a function of the quasi-periodic potential strength, and an intermediate phase between the thermal phase and the MBL phase shows up in both cases. This suggests a mechanism for the slow phase occurring in the AA model. The single-particle potential in the AA model is self-dual and thus fine-tuned. One basic effect of the interaction should be to break the self-duality, for example, in some mean-field sense. As a result, it is reasonable to argue that the single-particle potential is renormalized to a more generic one (without any self-duality) which features a single-particle mobility edge, making the model more GAA-like. The S phase could then be regarded as a many-body extension of an emergent single-particle mobility edge. A more systematic exploration of this proposed mechanism is left for future work.

%To summarize, we have studied the interacting AA model in detail using information scrambling and the entanglement structure of energy eigenstates. The results reveal an intermediate S phase between the thermal phase and the MBL phase which, when extended to the spinful interacting AA model, may be responsible for the slow dynamics recently observed in experiment.

\emph{Acknowledgment.}---This work is supported by Microsoft and Laboratory for Physical Sciences. The authors acknowledge the University of Maryland supercomputing resources made available for conducting the research reported in this paper. X.L. also acknowledges support from City University of Hong Kong (Project No. 9610428). B.G.S and S.X. acknowledge support for this work by the U.S. Department of Energy, Office of Science, Advanced Scientific Computing Research Quantum Algorithms Teams program as part of the QOALAS collaboration.

\bibliography{AA,scrambling,tensor}

\end{document}